\documentclass[journal=jpccck,manuscript=article]{achemso}
\usepackage{multirow}
\usepackage{chemformula} 
\usepackage[T1]{fontenc} 
\usepackage{hyperref}
\usepackage{comment}
\usepackage{amssymb}
\usepackage{siunitx}

\author{Kleuton A. L. Lima}
\affiliation[unicamp]
{Department of Applied Physics and Center for Computational Engineering and Sciences, State University of Campinas, Campinas, 13083-859, SP, Brazil}

\author{José A. S. Laranjeira}
\affiliation[unesp]
{Modeling and Molecular Simulation Group, São Paulo State University (UNESP), School of Sciences, Bauru, 17033-360, SP, Brazil}

\author{Nicolas F. Martins}
\affiliation[unesp]
{Modeling and Molecular Simulation Group, São Paulo State University (UNESP), School of Sciences, Bauru, 17033-360, SP, Brazil}

\author{Julio R. Sambrano}
\affiliation[unesp]
{Modeling and Molecular Simulation Group, São Paulo State University (UNESP), School of Sciences, Bauru, 17033-360, SP, Brazil}

\author{Alexandre C. Dias}
\affiliation[IFUnB]
{University of Bras\'ilia, Institute of Physics, 70910-900, Bras\'ilia, Federal District, Brazil.}

\author{Luiz A. Ribeiro Junior}
\affiliation[IFUnB]
{University of Bras\'ilia, Institute of Physics, 70910-900, Bras\'ilia, Federal District, Brazil.}
\alsoaffiliation[LCCMat]
{Computational Materials Laboratory, LCCMat, Institute of Physics, University of Bras\'ilia, 70910-900, Bras\'ilia, Federal District, Brazil.}
\email{ribeirojr@unb.br}

\author{Douglas S. Galvão}
\affiliation[unicamp]
{Department of Applied Physics and Center for Computational Engineering and Sciences, State University of Campinas, Campinas, 13083-859, SP, Brazil}

\title[Biphenylene]
  {First-Principles and Machine Learning Investigation of the Structural and Optoelectronic Properties of Dodecaphenylyne: A Novel Carbon Allotrope}


\begin{document}

\begin{tocentry}
\begin{center}
\includegraphics[width=0.8\linewidth]{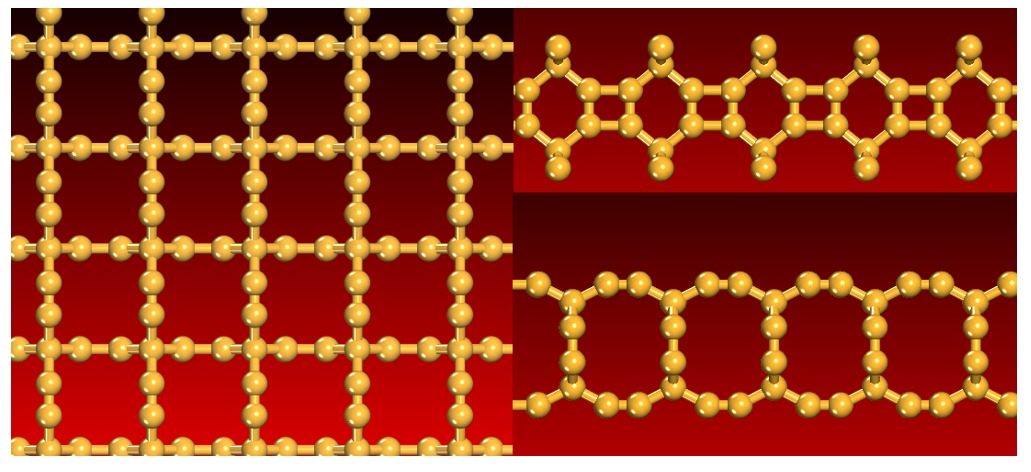}    
\end{center}
Dodecaphenylene: a semiconducting carbon allotrope with anisotropic properties.
\end{tocentry}

\begin{abstract}
\noindent We report the computational discovery and characterization of Dodecaphenylyne (DP), a novel carbon allotrope with a distinctive geometric arrangement. DP structural, thermodynamic, mechanical, electronic, and optical properties were evaluated using density functional theory and a machine learning interatomic potential trained explicitly for this material. The formation energy of –7.98 eV/atom indicates high thermodynamic stability, further supported by the absence of imaginary phonon modes and the preservation of structural integrity up to 1000 K in ab initio molecular dynamics simulations. Mechanical analysis reveals high in-plane stiffness with directional dependence: Young's modulus values of 469.09 GPa and 600.41 GPa along the x and y directions, respectively. Electronic band structure and projected density of states analyses confirm the DP semiconducting character. Calculations of carrier mobility using the deformation potential theory reveal pronounced anisotropy, with maximum values reaching up to $30.6 \times 10^4$ cm$^2$/V$\cdot$s (electrons, e) and $8.4 \times 10^4$ cm$^2$/V$\cdot$s (holes, h), much higher than the observed for other 2D materials. DP also exhibits anisotropic optical absorption in the visible and ultraviolet spectrum, highlighting its potential for optoelectronic applications. 
\end{abstract}

\section{Introduction}

Two-dimensional (2D) carbon-based materials have received considerable attention recently due to their exceptional physical properties and potential to revolutionize flat optoelectronic technologies~\cite{novoselov2004electric,geim2007rise}. The most important representative of this class of materials, graphene, exhibits unique properties, such as high electrical conductivity, high carrier mobility, and excellent mechanical strength, which make it an attractive candidate for applications ranging from transparent electrodes to high-frequency transistors~\cite{castro2009electronic}. However, graphene's intrinsic zero bandgap remains a significant limitation for its use in digital and optoelectronic devices that require well-defined on/off states and efficient light-matter interactions~\cite{schwierz2010graphene}.

This limitation has stimulated the search to discover novel graphene-like allotropes and related 2D carbon networks with tunable electronic behavior \cite{enyashin2011graphene,hirsch2010era}. Numerous theoretical and experimental investigations have proposed new carbon architectures, such as graphyne \cite{desyatkin2022scalable}, reacted graphyne (5,6,9-Ringene) \cite{aliev2025planar}, graphdiyne \cite{gao2019graphdiyne}, phagraphene \cite{wang2015phagraphene}, DOTT-Carbon \cite{lima2025first}, PolyPyGY \cite{LIMA2025116099}, Holey-Graphyne \cite{liu2022constructing}, Irida-Graphene \cite{junior2023irida}, and biphenylene network \cite{fan2021biphenylene}. Despite this progress, many materials have a metallic-like electronic structure.

Addressing this limitation, we propose here a 2D carbon allotrope, hereafter referred to as dodecaphenylyne (DP), which combines structural features from linear carbynes and the recently synthesized biphenylene network~\cite{fan2021biphenylene}. The design strategy exploits the structural rigidity and planarity of sp- and sp$^2$-hybridized carbon frameworks to engineer a periodic lattice composed of interconnected four-, six-, and twelve-membered carbon rings. 

We have employed a combination of density functional theory (DFT) and a machine learning-based interatomic potential (MLIP) within the framework of classical reactive molecular dynamics (MD) simulations. Its structural integrity is through formation energy, phonon dispersion analyses, and thermal stability via \textit{ab initio} molecular dynamics (AIMD) simulations. Mechanical properties, including Young's modulus values, are obtained using the trained MLIP. We also explore the electronic band structure, projected density of states, and effective masses of carriers. Finally, we analyzed the optical absorption spectra and demonstrated the material's anisotropic behavior in the visible and ultraviolet regions. 

\section{Methodology}

DFT calculations were carried out using the Vienna Ab initio Simulation Package (VASP)~\cite{kresse1993ab,kresse1996efficiency}, employing the projector augmented wave (PAW) method. For structural optimizations and total energy calculations, the Perdew–Burke–Ernzerhof (PBE) exchange-correlation functional within the generalized gradient approximation (GGA) was employed~\cite{perdew1996generalized}. To improve the accuracy of electronic property predictions, particularly the description of band dispersion and localization effects, additional calculations were performed using the hybrid functional HSE06~\cite{heyd2003hybrid}. A plane-wave energy cutoff of 450~eV was used, and the Brillouin zone was sampled with a $\Gamma$-centered Monkhorst-Pack mesh of $10\times10\times1$.

Geometric optimizations were performed under periodic boundary conditions until the total energy and residual forces converged to less than $1.0 \times 10^{-5}$~eV and $1.0 \times 10^{-3}$~eV/\AA, respectively. The stress tensor was monitored to ensure that the in-plane pressure remained below 0.01 GPa. A vacuum region of 20~\AA~was introduced along the out-of-plane ($z$) direction to avoid spurious interlayer interactions.

Phonon dispersion curves were calculated using density functional perturbation theory (DFPT)~\cite{baroni2001phonons}, as implemented in the PHONOPY package~\cite{togo2015first}, to evaluate the DP dynamical stability. In addition, AIMD simulations were carried out at 1000~K for 5~ps with a 1~fs timestep using a Nosé-Hoover thermostat~\cite{nose1984unified} in the canonical (NVT) ensemble. These simulations aimed to probe the thermal robustness of the structure by monitoring atomic displacements and bond changes.

The electronic band structure and projected density of states (PDOS) were computed using a $10\times10\times1$ and $20\times20\times1$ $k$-point mesh, respectively. The electron localization function (ELF) and PDOS were analyzed to provide insights into the nature of bonding and charge delocalization. Effective masses for electrons and holes were estimated from the curvature of the band edges along high-symmetry directions. 

For the excitonic and optical properties, we first obtained a maximally localized Wannier function tight-binding (MLWF-TB) Hamiltonian directly from the HSE06 calculation, using the Wannier90 code,\cite{Arash_685_2008}, considering C $ s$- and $ p$-orbital projections. The linear optical response was obtained using the independent particle approximation (IPA) and through the solution of the Bethe-Salpeter equation (BSE)\cite{Salpeter_1232_1951} to account for the excitonic effects. The electron-hole Coulomb interaction in BSE was modeled using the 2D Coulomb Truncated Potential (V2DT).\cite{Rozzi_205119_2006} Both calculations were carried out using the WanTiBEXOS code,\cite{Dias_108636_2022} using the three highest valence bands and the lowest conduction band, which is sufficient to describe the linear optical response in the solar emission range, i.e. \SIrange{0}{4}{\electronvolt}, we also used a smearing of \SI{0.05}{\electronvolt} for the calculation of real and imaginary dielectric functions.

To assess the DP mechanical response, classical molecular dynamics simulations were performed using the LAMMPS package~\cite{plimpton1995fast}. A machine-learned moment tensor potential (MTP)~\cite{shapeev2020elinvar,novikov2020mlip} was explicitly developed for this material using the MLIP framework~\cite{podryabinkin2017active}. Training data for the potential were derived from DFT-based AIMD simulations on both relaxed and strained supercells across different temperatures. The velocity Verlet algorithm was used to integrate Newton’s equations of motion with a time step of 0.05 fs. The systems were equilibrated under an isothermal-isobaric (NPT) ensemble for 100~ps before applying tensile strain (uniaxial and biaxial).

Mechanical testing simulations were performed at 300~K under constant engineering strain rates of $10^{-6}$~fs$^{-1}$. Stress-strain curves were obtained from these simulations, and elastic moduli and Poisson’s ratios were calculated in both in-plane directions. The use of MLIPs in evaluating the mechanical performance of low-dimensional carbon materials has been widely validated and has shown excellent agreement with first-principles results in prior studies~\cite{mortazavi2023machine,mortazavi2023atomistic,mortazavi2025exploring,mortazavi2020efficient,mortazavi2023electronic,mortazavi2024goldene,mortazavi2022anisotropic,lima2024th,lima2025first,LIMA2025116099}.

\section{Results}

\subsection{Structural Properties}

\begin{figure}[!htb]
    \centering
    \includegraphics[width=\textwidth]{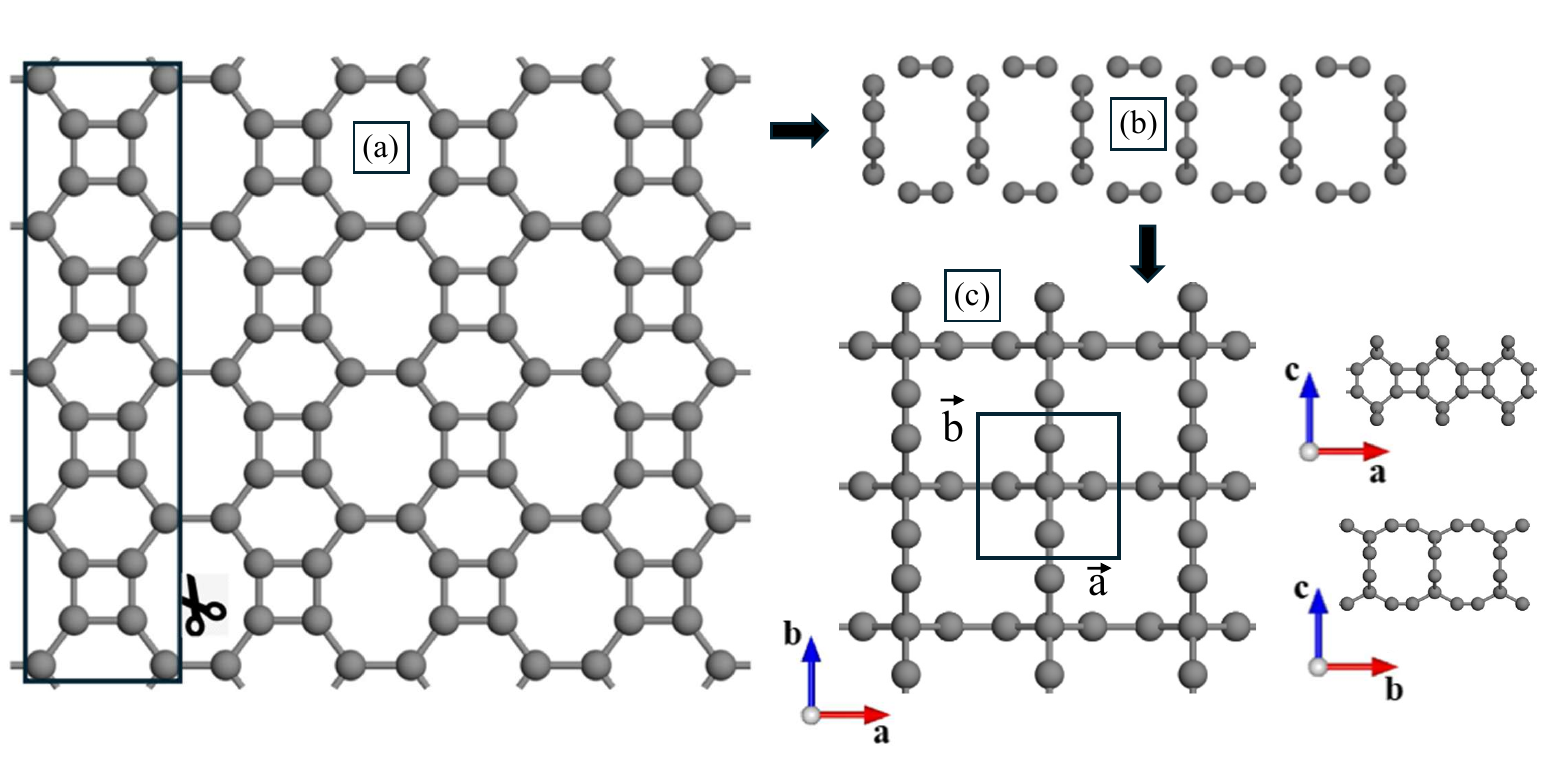}
    \caption{Optimized DP structure. (a) The top view shows the BPN sheet, the (b) side view, and the (c) top view of DP, with the highlighted unit cell. Carbon atoms are shown as grey spheres. The crystal structure is orthorhombic, with lattice parameters $a = 3.92$~\AA and $b = 3.85$~\AA. The rightmost panel illustrates the two distinct side views of DP.}
    \label{fig:structure}
\end{figure}

Figure~\ref{fig:structure} illustrates the atomic arrangement of the newly proposed DP. Figure~\ref{fig:structure}(a) shows the top view of the optimized biphenylene network (BPN)~\cite{fan2021biphenylene} sheet (black rectangle with a scissor). The DP structure consists of interconnected square and hexagonal rings, forming a grid-like pattern distinct from the honeycomb lattice of graphene or the acetylenic linkages of graphyne (see Figures~\ref{fig:structure}(b,c)).

Figures \ref{fig:structure}(b) and \ref{fig:structure}(c) show the side views along the $a$- and $b$-axes, respectively, showing negligible out-of-plane distortions. DP belongs to the orthorhombic space group PMMM (D\textsubscript{2h}-1) with lattice parameters $a = 3.92$~\AA and $b = 3.85$~\AA. All lattice angles are 90$^{\circ}$, consistent with the orthorhombic symmetry, and the structure exhibits no imaginary phonon modes, confirming its dynamical stability. The vertical separation between the top and bottom atomic positions defines a thickness of approximately 4.61~\AA. The supplementary material contains the DP structure CIF file.

Compared to graphene, which consists purely of hexagonal rings and exhibits Dirac-cone behavior symmetry~\cite{castro2009electronic}, DP breaks hexagonal symmetry by incorporating four-, six-, and twelve-membered rings, with the first two related to the BPN motifs~\cite{fan2021biphenylene}. The presence of highly distorted ring geometries and a broad distribution of C–C bond length values, varying from 1.22 to 1.56~\AA, indicates a mixed hybridization character in the lattice. In contrast to graphyne and graphdiyne, which exhibit an association of sp and sp$^2$ hybridizations (linked by linear acetylenic chains~\cite{malko2012graphyne}), DP combines sp$^2$ and sp$^1$ hybridized carbon atoms in a contiguous 2D framework.

An important DP structural feature is the variation in carbon–carbon bond lengths, which range from 1.22 to 1.56~\AA\ across the lattice. This broad distribution arises from square and hexagonal ring motifs, leading to diverse local environments and bond geometries. Shorter bonds (around 1.22~\AA) are typically associated with stronger $\sigma$ or delocalized $\pi$ bonding, while longer bonds (up to 1.56~\AA) suggest the presence of weaker, possibly strained, or sterically constrained interactions. As discussed later, such bond length variation may induce electronic localization effects and contribute to the anisotropy observed in electronic and mechanical properties. 

The DP formation energy (\(E_{\text{f}}\)) was found to be $-7.98$~eV/atom, a value comparable to many other theoretically predicted and stable 2D carbon allotropes \cite{cavalheiro2024can}, and to that of other experimentally realized structures such as BPN \cite{niu2023unveiling} and Holey-Graphyne \cite{mortazavi2023electronic}. This result indicates a thermodynamically stable configuration, suggesting potential feasibility for synthesis under suitable experimental conditions, such as surface-assisted polymerization or chemical vapor deposition, which were used to obtain other carbon allotropes.

\begin{figure}[!htb]
    \centering
    \includegraphics[width=\textwidth]{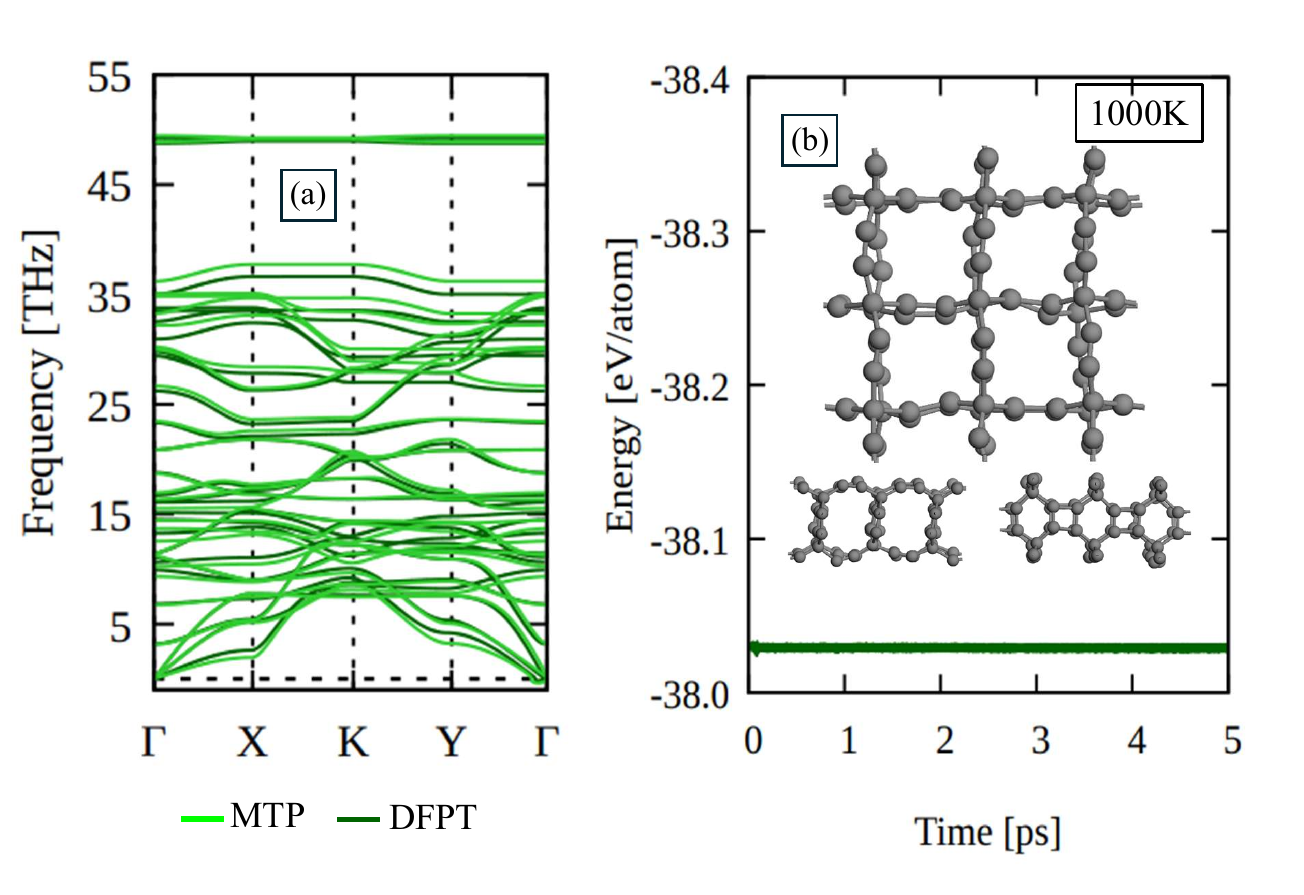}
    \caption{(a) DP phonon dispersion computed using DFPT and MTP. (b) Time evolution of total energy per atom during AIMD simulation at 1000 K for 5 ps.}
    \label{fig:stability}
\end{figure}

Figure~\ref{fig:stability}(a) shows the DP phonon dispersion calculated using two independent approaches: density functional perturbation theory (DFPT) and the machine learning-based moment tensor potential (MTP). The excellent agreement between both methods across the entire Brillouin zone confirms the accuracy and transferability of the trained potential for vibrational analysis. Importantly, no imaginary frequencies are observed in either dispersion, indicating that the structure is dynamically stable and resides at a minimum of the potential energy surface.

The phonon branches extend up to approximately 48~THz, consistent with strong covalent bonding between carbon atoms and comparable to the high-frequency modes observed in graphene and other sp$^2$-based 2D carbon systems. The relatively well-separated optical and acoustic branches suggest well-defined phonon group velocities, which may affect thermal transport properties.

To further investigate the structural integrity of DP under finite-temperature conditions, AIMD simulations were performed at 1000 K for a duration of 5 ps. Figure~\ref{fig:stability}(b) presents the time evolution of the total energy per atom during the simulation. The energy remains stable with minimal fluctuations, and no abrupt jumps or drops are observed, which indicates that no phase transition, bond breaking, or structural collapse occurs throughout the trajectory. Snapshots of the final atomic configuration, shown as insets in Figure~\ref{fig:stability}(b), confirm that the original topology is fully preserved after thermal agitation. 

\subsection{Electronic Properties}

\begin{figure}[!htb]
    \centering
    \includegraphics[width=\textwidth]{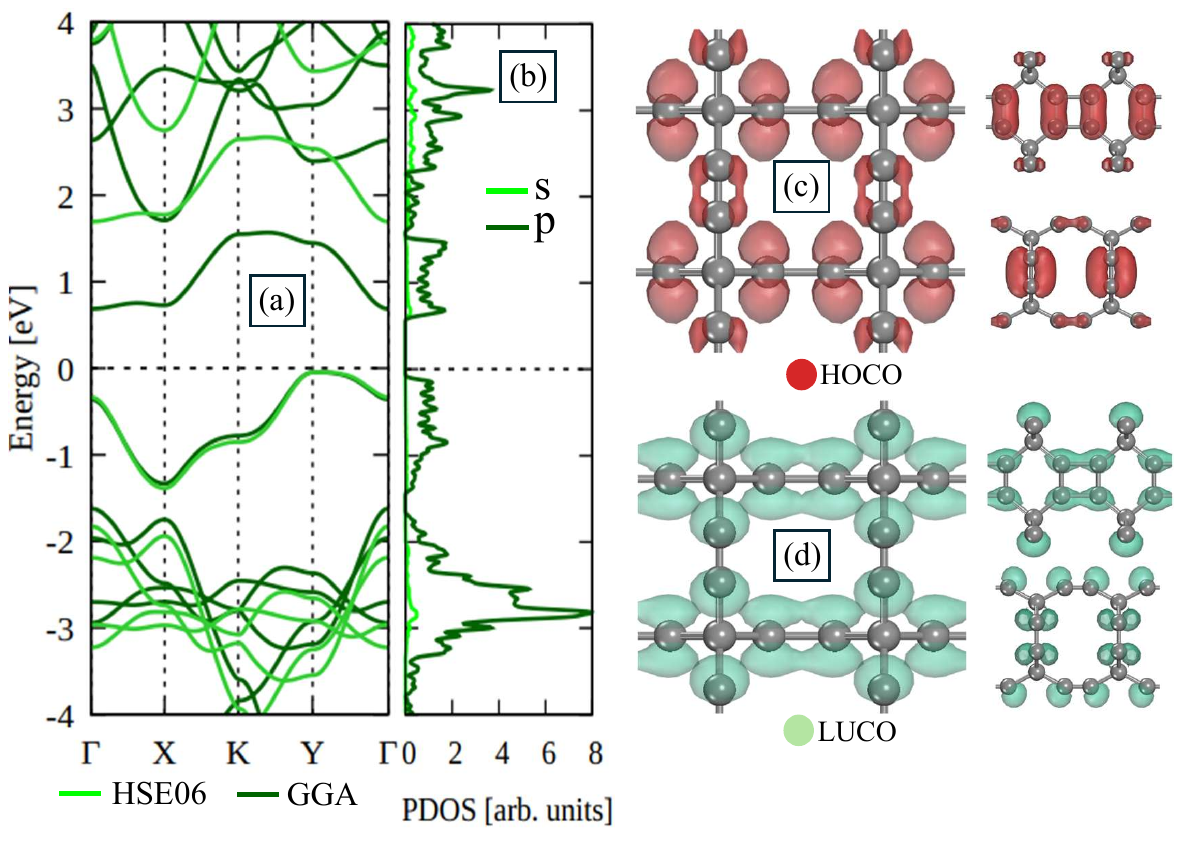}
    \caption{(a) Electronic band structure of DP calculated using PBE and HSE06. (b) PDOS for $s$ and $p$ orbitals. (c,d) Real-space isosurfaces of the highest occupied crystalline orbital (HOCO, red) and the lowest unoccupied crystalline orbital (LUCO, green).}
    \label{fig:electronic}
\end{figure}

Figure~\ref{fig:electronic}(a) displays the electronic band structure of DP calculated using the PBE and HSE06 methods. While the PBE functional tends to underestimate the band gap, the HSE06 results reveal that DP is a semiconductor. The band gap is indirect, with the valence band maximum (VBM) located between the X and K points and the conduction band minimum (CBM) positioned near the Y point. The band gap obtained with HSE06 is approximately 1.73~eV, placing the material within the ideal range for photovoltaic applications.

The PDOS shown in Figure~\ref{fig:electronic}(b) indicates that the electronic states near the Fermi level are dominated by carbon $p$-orbitals, as expected for a conjugated sp$^2$-hybridized carbon network. The contribution of $s$-orbitals is relatively small and primarily from deeper valence states. The sharp features in the PDOS near the band edges suggest well-defined electronic states, which may result in favorable charge carrier mobilities. 

Figures~\ref{fig:electronic}(c) and (d) show the spatial distribution of the highest occupied crystalline orbital (HOCO, red) and the lowest unoccupied crystalline orbital (LUCO, green). The HOCO is delocalized along the backbone of the carbon framework, particularly around the six-membered rings, indicating an extended conjugation in the valence states. LUCO appears more localized along the four-membered ring regions and acetylenic bonds. 

\begin{figure}[H]
    \centering
    \includegraphics[width=\textwidth]{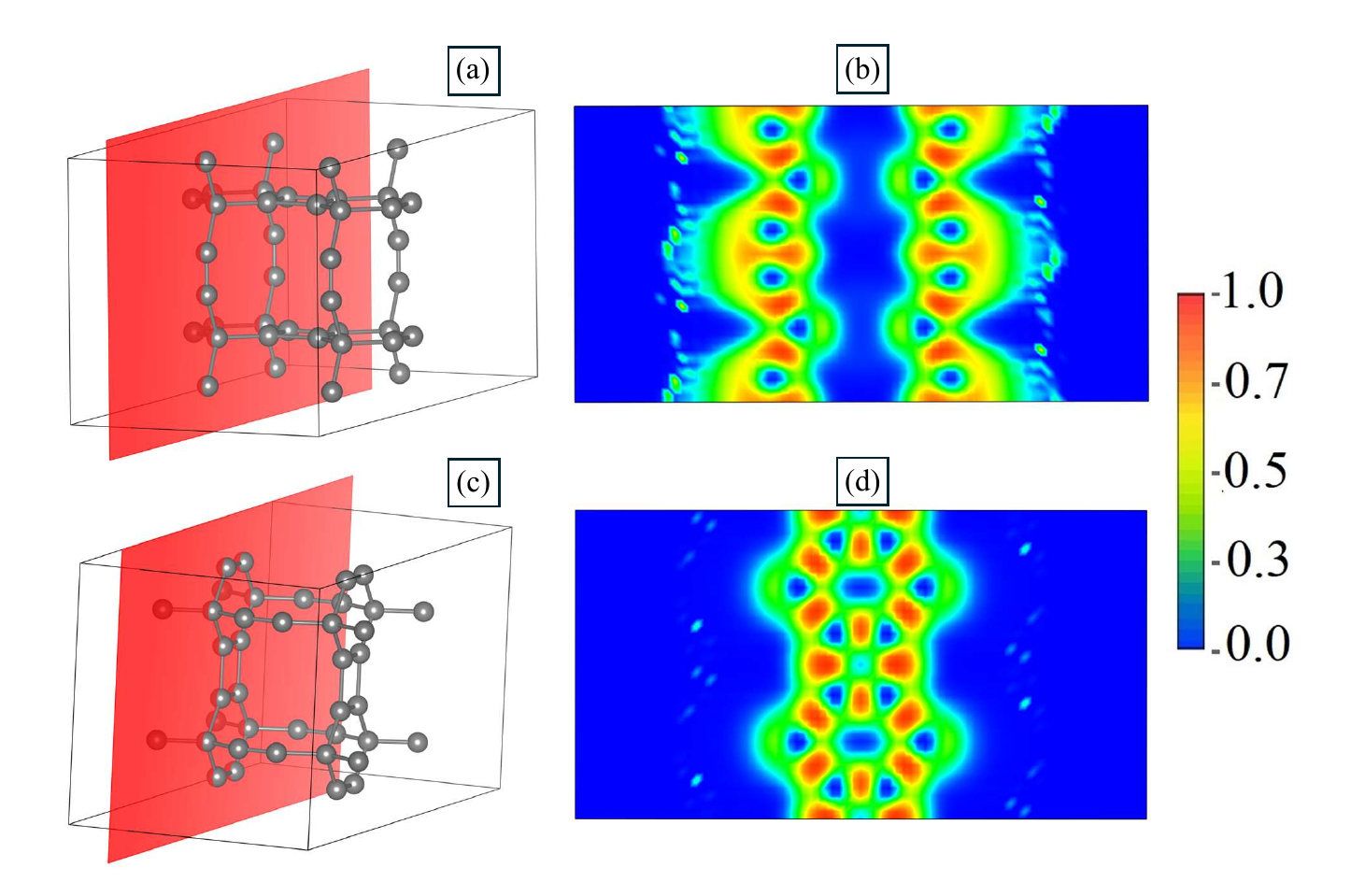}
    \caption{(a,c) Structural planes selected for ELF mapping. (b,d) Electron localization function (ELF) projected onto the respective planes. High ELF values (red) indicate strong electron localization and covalent bonding regions, while low values (blue) correspond to delocalized electron density.}
    \label{fig:elf}
\end{figure}

Figure~\ref{fig:elf} illustrates the DP electron localization function (ELF), providing insight into the nature of chemical bonding and charge localization within the structure. Figures \ref{fig:elf}(a) and \ref{fig:elf}(c) depict the planes selected for analysis, while Figures \ref{fig:elf}(b) and \ref{fig:elf}(d) show the corresponding ELF maps projected onto those planes.

ELF values range from 0 (blue, delocalized) to 1 (red, highly localized). The ELF map in Figure \ref{fig:elf}(b), corresponding to the central vertical plane (Figure~\ref{fig:elf}(a)]), reveals high localization regions (ELF > 0.7) surrounding the majority of C–C bonds, indicating strong covalent bonding consistent with sp$^2$ hybridization. These high ELF values are presented along the edges of the hexagonal rings, where $\pi$-bonding character dominates.

In contrast, the ELF slice shown in Figure \ref{fig:elf}(d), taken from an off-center diagonal plane (Figure~\ref{fig:elf}(c)), exhibits a more complex pattern of localization. In addition to the expected bonding regions, localized charge density is observed in slightly elevated zones away from the atomic centers. This feature reflects the partial sp$^2$ hybridization in some carbon atoms, especially those involved in strained four-membered rings.

\subsection{Effective Mass and Carrier Mobility}

To investigate DP charge transport properties, we calculated the effective masses and mobilities of electrons and holes under uniaxial strain along the $x$ and $y$ directions. The carrier effective mass $m^*$ is obtained from the curvature of the energy dispersion near the band extrema using the standard parabolic approximation:

\begin{equation}
m^* = \hbar^2 \left( \frac{d^2E}{dk^2} \right)^{-1}
\end{equation}

\noindent where $E(k)$ is the energy of the conduction or valence band and $\hbar$ is the reduced Planck constant. 

\begin{figure}[!htb]
    \centering
    \includegraphics[width=\textwidth]{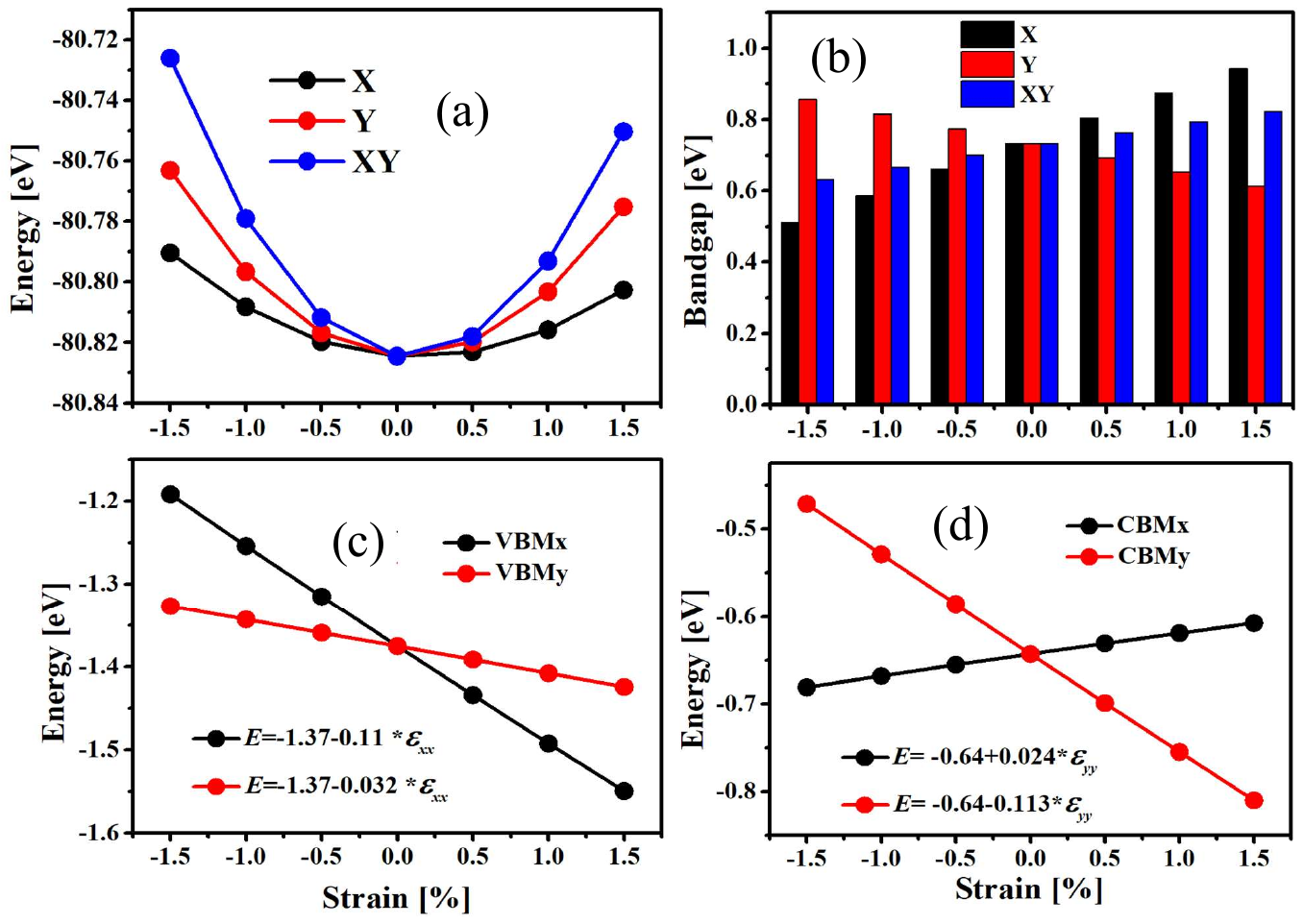}
    \caption{(a) DP total energy variation under uniaxial strain along $x$, $y$, and $xy$ directions. (b) Electronic band gap as a function of strain. (c) Energy shift of the valence band maximum (VBM) under strain along $x$ and $y$ directions. (d) Energy shift of the conduction band minimum (CBM) under strain. Slopes are used to obtain the deformation potential constants.}
    \label{fig:mobility}
\end{figure}

Figure~\ref{fig:mobility}(a) shows the variation of the total energy under the uniaxial strain along the $x$ and $y$ directions, as well as under the biaxial ($xy$) strain. The parabolic profiles confirm the elastic nature of the deformation within the range of $\pm1.5\%$. Figure~\ref{fig:mobility}(b) shows the corresponding electronic band gap evolution, where a strong dependence on both the strain direction and magnitude is observed. Under uniaxial strain along the $x$-direction (black bars), the band gap increases with tensile strain and decreases under compression. 

In contrast, when the strain is applied along the $y$-direction (red bars), the band gap increases under compressive strain and decreases with stretching. A symmetric response is observed for biaxial strain (blue bars): the band gap increases under tensile strain and decreases under compressive strain. This directional asymmetry highlights the DP's anisotropic electronic structure and is consistent with the orbital localization and hybridization patterns revealed by ELF and band-edge analyses.

Figures ~\ref{fig:mobility}(c,d) provide linear fits to the energy shifts of the valence band maximum (VBM) and conduction band minimum (CBM) under strain, from which the deformation potential constants $E_1$ are obtained. These are needed to estimate the charge carrier mobility $\mu$ using the deformation potential theory developed by Bardeen and Shockley:

\begin{equation}
\mu = \frac{2e\hbar^3C_{2D}}{3k_BT |E_1|^2 (m^*)^2}
\end{equation}

\noindent where $C_{2D}$ is the in-plane stiffness, $k_B$ is the Boltzmann constant, and $T$ is the temperature (set to 300~K in this study). The values of $C_{2D}$ were obtained from the energy–strain relations using:

\begin{equation}
C_{2D} = \frac{1}{S_0} \frac{d^2E}{d\varepsilon^2}
\end{equation}

\noindent where $S_0$ is the equilibrium area of the unit cell, and $\varepsilon$ is the applied strain. All the fitted constants and resulting effective masses and mobilities are summarized in Table~\ref{mobility}.

\begin{table*}[!htbp]
\centering
\caption{Calculated effective mass $m_i^*$, average effective mass $m_d$, in-plane stiffness $C_{2D}$, deformation potential constant $E_1$, and DP charge carrier mobility $\mu$ at $T=300$~K. Here, $m_0$ is the mass of a free electron.}
\label{mobility}
\begin{tabular}{|c|c|c|c|c|c|c|}
\hline
Direction & Carrier & $m_i^*$ ($m_0$) & $m_d$ ($m_0$) & $C_{2D}$ (eV/\AA$^2$) & $E_1$ (eV) & $\mu$ ($10^4$ cm$^2$/V$\cdot$s) \\
\hline
$x$ & $e$ & 0.857 & 2.75 & 14.62 & 0.024 & 30.6 \\
$x$ & $h$ & 0.737 & 3.86 & 14.62 & -0.118 & 8.4 \\
\hline
$y$ & $e$ & 1.105 & 2.75 & 19.90 & -0.113 & 11.6 \\
$y$ & $h$ & 2.94 & 3.86 & 19.90 & -0.032 & 5.6 \\
\hline
\end{tabular}
\end{table*}

The calculated results reveal that DP exhibits anisotropic properties. Electron mobility is higher along the $x$-direction, reaching values above $30.6 \times 10^4$~cm$^2$/V$\cdot$s, while hole mobility is higher along the same direction, with $8.4 \times 10^4$~cm$^2$/V$\cdot$s. Along the $y$-direction, electron mobility reaches $11.6 \times 10^4$~cm$^2$/V$\cdot$s, while hole mobility drops to $5.6 \times 10^4$~cm$^2$/V$\cdot$s. This anisotropy can be attributed to the directional dependence of orbital overlap and band edge curvature. 

Our findings point to DP mobilities values that surpass the reported values for different 2D materials, such as Tetrahexcarbon ($1.46 \times 10^4$~cm$^2$/V$\cdot$s)\cite{peng2020enhanced}, Orth-C$_16$ ($8.30 \times 10^4$~cm$^2$/V$\cdot$s)\cite{liu2024two}, black phosphorene ($2.60 \times 10^4$~cm$^2$/V$\cdot$s)\cite{qiao2014high}, and MoS$_2$ ($0.05 \times 10^4$~cm$^2$/V$\cdot$s)\cite{radisavljevic2011single}. 

\subsection{Excitonic and Optical Properties}

\begin{figure}[!h]
    \centering
    \includegraphics[width=0.6\linewidth]{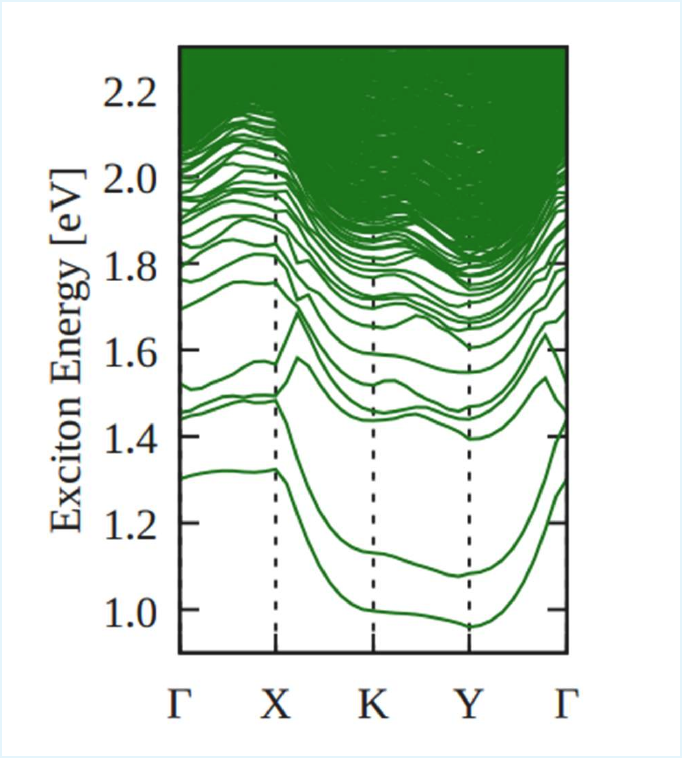}
    \caption{DP exciton band structure.}
    \label{fig:exc_bands}
\end{figure}

Excitonic effects are fundamental for an accurate description of the linear optical response in 2D materials.\cite{Dias_3265_2021,cavalheiro2024can} The excitonic band structure is shown in Fig.~\ref{fig:exc_bands}, revealing that the excitonic ground state is indirect, i.e., the electron and hole are located at different \textbf{k}-points, as expected due to the indirect nature of the fundamental electronic band gap. This indirect excitonic ground state is located in the vicinity of $Y$ high symmetry point, at \SI{0.96}{\electronvolt} and the direct excitonic ground state (located at $\Gamma$) at \SI{1.30}{\electronvolt}. 

When comparing the exciton ground state with the fundamental band gap (\SI{1.74}{\electronvolt}), we can estimate an exciton binding energy of \SI{779}{\milli\electronvolt}, with is higher than the values expected for 2D carbon allotropes \cite{cavalheiro2024can} and other 2D materials,\cite{Moujaes_111573_2023,Guassi_62094_2024,Huacarpuma__2025} where the exciton binding energy is found around \SIrange{100}{500}{\milli\electronvolt}.\cite{Dias_3265_2021,cavalheiro2024can,Huacarpuma__2025} The presence of an indirect excitonic ground state suggests the possibility of phonon-assisted optical transition, with photon excitations lower than the direct excitonic ground state. 

\begin{figure}[!h]
    \centering
    \includegraphics[width=1\linewidth]{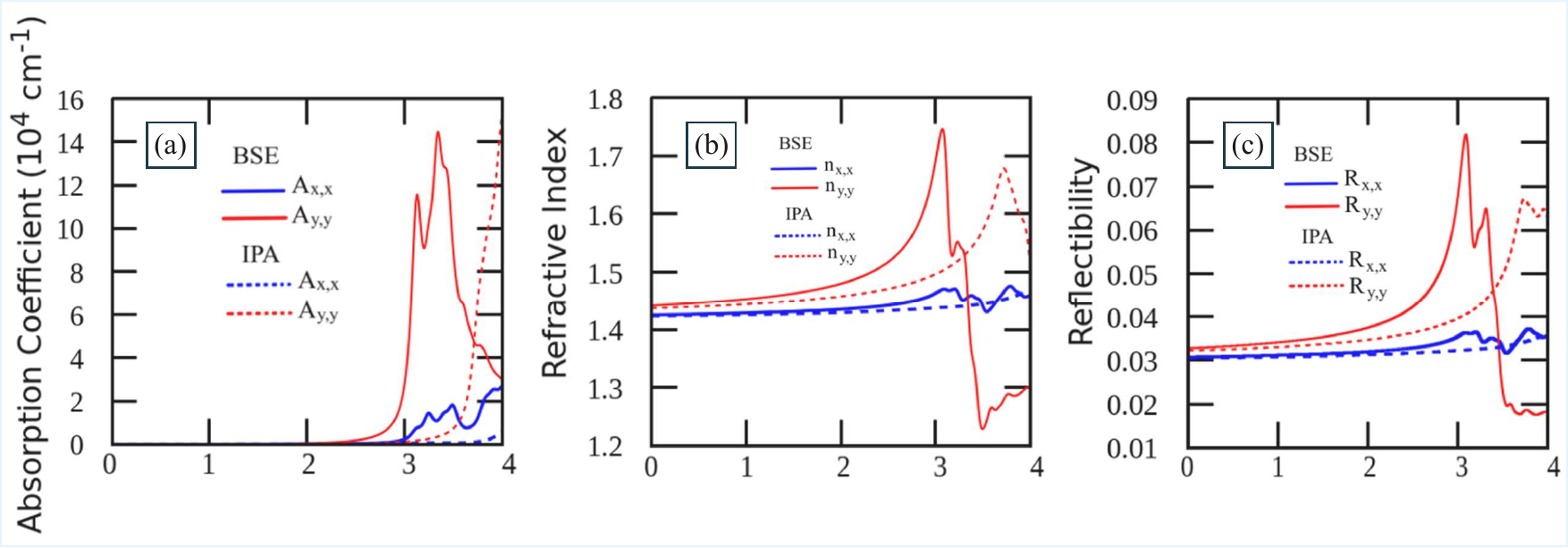}
    \caption{DP optical properties, at BSE (solid curves) and IPA (dashed curves) levels: (a) Absorption Coefficient, (b) Refractive Index, (c) Reflectibility; considering a linear light polarization at $\hat{x}$ (blue curves) and $\hat{y}$ (red curves) directions.}
    \label{fig:optics}
\end{figure}

The linear optical response is illustrated in Fig.~\ref{fig:optics} through the absorption coefficient, refractive index, and reflectivity, obtained using the IPA and BSE frameworks, considering linear light polarization along the $\hat{x}$ and $\hat{y}$ directions. 

From the absorption coefficient, in Fig.~\ref{fig:optics}(a), we observe that the monolayer is highly anisotropic, showing an optical response eight times larger for light polarization for the $\hat{y}$ direction.

We only observe significant optical response closer to \SI{3}{\electronvolt}, which could be explained by the lower concentration of electron and hole single particle states closer to the Fermi level, as shown in the electronic band structure and density of states (Fig.\ref{fig:electronic}). These results, combined with the electron and hole orbital symmetries of the states below \SI{3}{\electronvolt}, result in a small oscillator force, i.e., a small optical transition probability. 

This system has a higher absorption coefficient in the visible and ultraviolet regions of the spectrum. Due to the larger exciton binding energy of \SI{779}{\milli\electronvolt}, the red shift in the absorption curve is easily observed and significant, indicating strong excitonic effects. At the IPA level, it exhibits considerable absorption only in the ultraviolet region.

The refractive index and reflectivity, as shown in Figs.~\ref{fig:optics}(b) and (c), exhibit the same trend, with significant values at the end of the visible and ultraviolet regions, showing a higher degree of anisotropy. At the BSE level, the refractive index reaches a peak of \SI{1.7}{} at \SI{3}{\electronvolt}, while at the IPA level, this peak has a smaller intensity. It is located around \SI{3.7}{\electronvolt}, both results considering a linear light at the $\hat{y}$ direction. For the $\hat{x}$ direction, the refractive index is between \SIrange{1.4}{1.5}{} independent of the photon excitation. The reflectibility follows the same trend of the refractive index, reaching a maximum of \SI{8}{\percent} at BSE level with a $\hat{y}$ polarized light, for photon excitations closer to \SI{3}{\electronvolt}.

\subsection{Mechanical Properties}

\begin{figure}[!htb]
    \centering
    \includegraphics[width=0.7\textwidth]{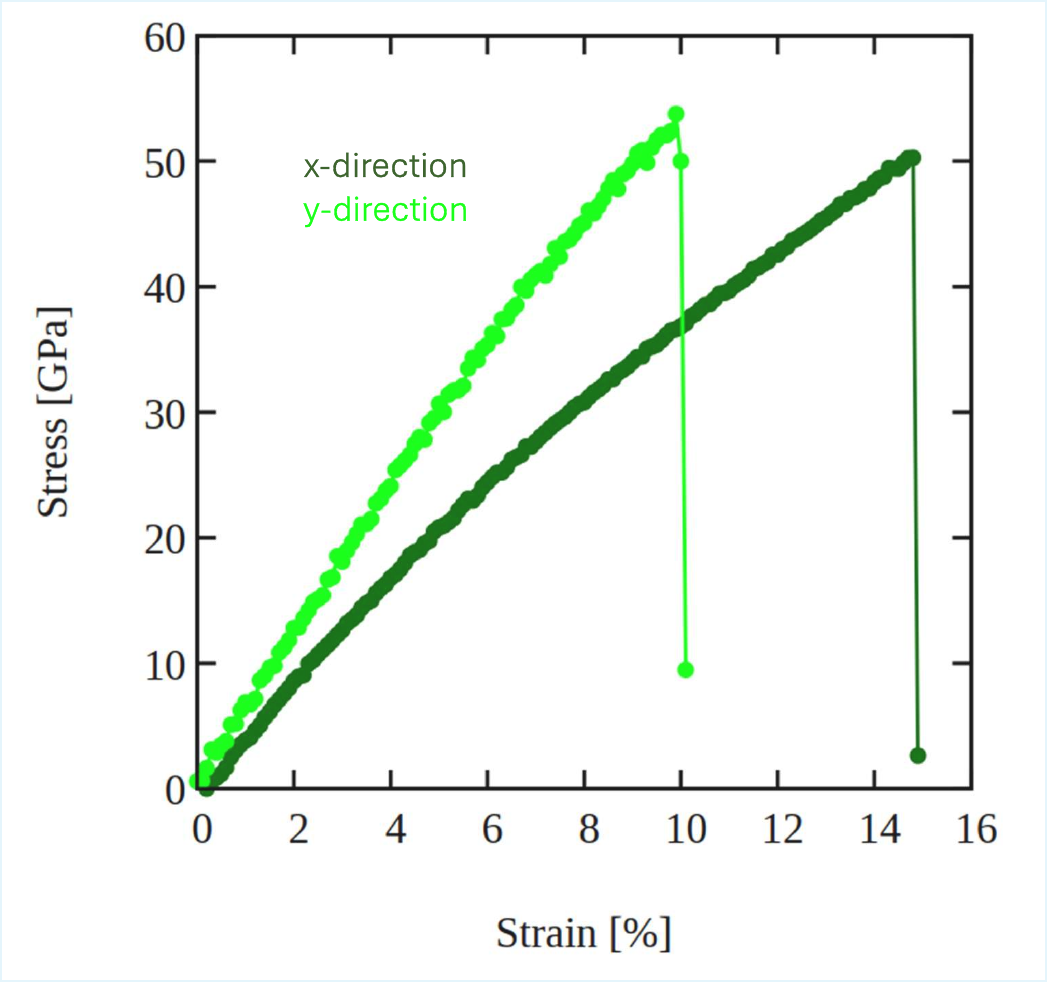}
    \caption{DP uniaxial stress-strain curves for the $x$- and $y$-directions. The slopes in the elastic regime (up to 1\% strain) yield Young's modulus values of 469.1~GPa and 600.4~GPa, respectively.}
    \label{fig:stressstrain}
\end{figure}

Figure~\ref{fig:stressstrain} shows the DP uniaxial stress-strain curves along the $x$- and $y$-directions. The curves were obtained from reactive classical MD simulations based on the MLIP framework and trained with AIMD data. Both directions exhibit a well-defined elastic region followed by nonlinear behavior and eventual structural failure, revealing a significant anisotropy in the material's mechanical response.

Young's modulus values are estimated from the slope of the stress-strain curve in the linear regime up to a strain of 1 \%. The obtained values are approximately 469.1 GPa along the $x$-direction and 600.4 GPa along the $y$-direction, indicating a higher stiffness along the $y$-axis. Importantly, these values are considerably smaller than those for graphene, about 1 TPa \cite{scarpa2009effective,lee2012estimation}. This anisotropy is consistent with the underlying atomic arrangement, where the denser bonding network along the $y$ direction provides greater resistance to deformation.

The curves provide the ultimate tensile stress and the corresponding fracture strain for each direction. Along the $x$-direction (light green curve), the material withstands a maximum stress of approximately 51.4~GPa at 9.9\% strain. In contrast, along the $y$-direction (dark green curve), the material sustains up to 52.9~GPa at a higher strain of 14.8\%. This difference illustrates that although the material is stiffer along $y$, it is also more ductile in that direction, sustaining larger deformations before failure.

A sharp drop is observed after the peak stress is reached, indicating the onset of fracture. The post-peak fluctuations in the stress-strain response suggest bond breaking and atomic rearrangements during crack propagation. The observed failure pattern indicates a brittle-to-quasi-brittle transition characterized by distinct fracture mechanisms that depend on the direction of loading. Along the $x$ direction, the rapid stress drop indicates brittle rupture with minimal plasticity, whereas the gradual decrease along $y$ suggests local plastic deformation preceding complete failure.

\begin{figure}[!htb]
    \centering
    \includegraphics[width=\textwidth]{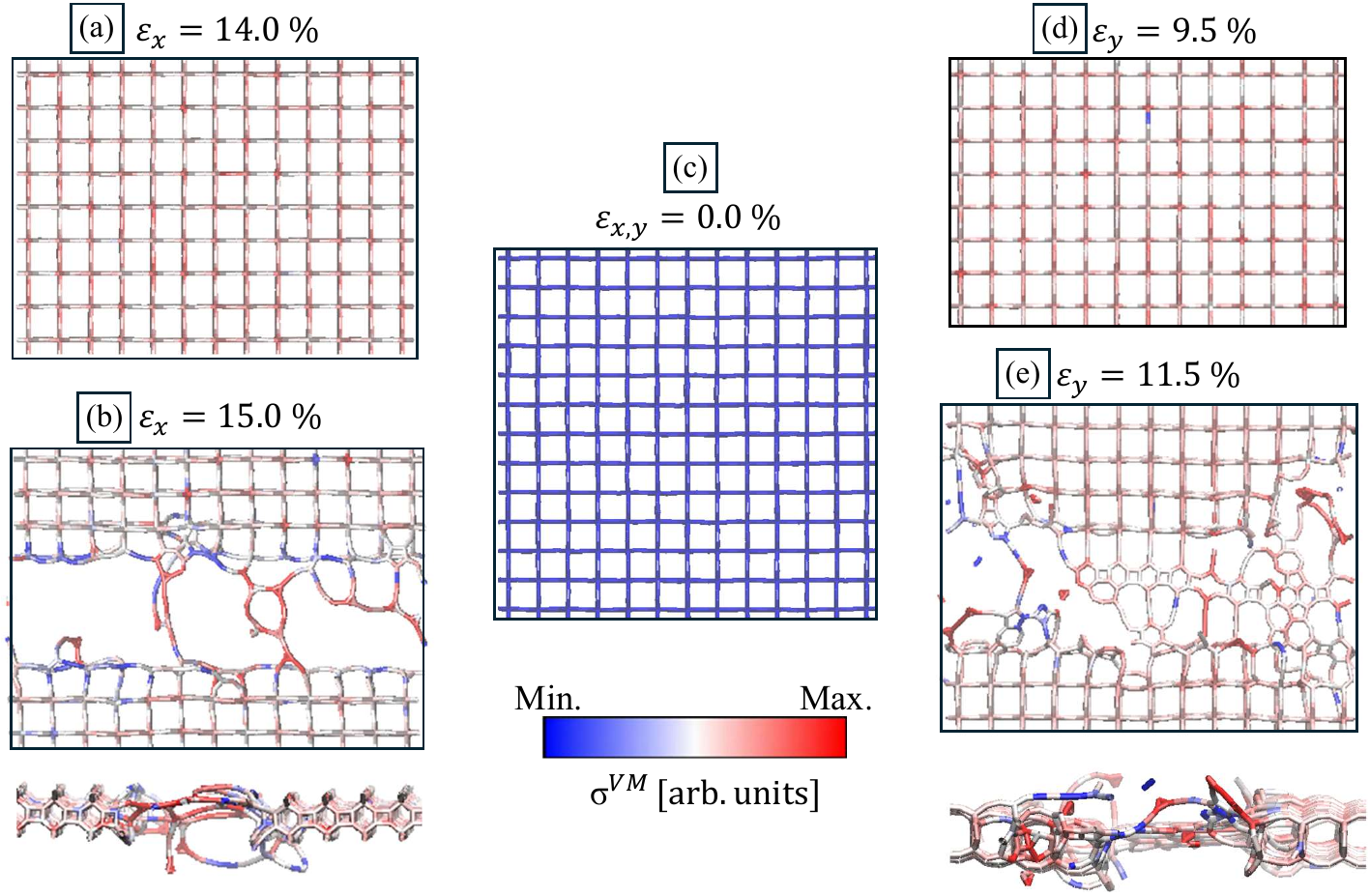}
    \caption{Representative snapshots from molecular dynamics simulations showing DP atomic structures at different strain levels. (a,b) Uniaxial strain along $x$ at 14.0\% and 15.0\%; (d,e) uniaxial strain along $y$ at 9.5\% and 11.5\%. (c) Equilibrium configuration at 0\% strain. Atoms are colored according to von Mises stress values. Side views highlight fracture morphology and structural deformations.}
    \label{fig:fracture}
\end{figure}

Finally, we discuss the DP fracture patterns and failure mechanisms. Figure~\ref{fig:fracture} presents representative MD snapshots under uniaxial strain, colored according to the local von Mises stress distribution \cite{felix2020mechanical}. The central panel (Figure \ref{fig:fracture}(c)) displays the undeformed equilibrium structure ($\varepsilon = 0\%$), characterized by a uniform and minimal stress field, which serves as the reference configuration. Figures \ref{fig:fracture}(a) and \ref{fig:fracture}(b) correspond to deformation along the $x$-direction at 14.0\% and 15.0\% strain, respectively, while Figures \ref{fig:fracture}(d) and \ref{fig:fracture}(e) show deformations along the $y$-direction at 9.5\% and 11.5\% strain, respectively. The bottom panels in Figures \ref{fig:fracture}(b) and \ref{fig:fracture}(e) show side views of the fractured structures.

At 14.0\% strain along the $x$ direction (Figure~\ref{fig:fracture}(a)), the stress distribution remains relatively homogeneous, with most atoms experiencing moderate von Mises stress. However, just before structural failure (15.0\%, Figure~\ref{fig:fracture}(b)), high-level stress accumulates near the four-membered rings, leading to the nucleation of cracks. The side view in Figure~\ref{fig:fracture}(b) reveals significant out-of-plane deformations and structural tearing, consistent with the sharp drop in the stress-strain curve (Figure~\ref{fig:stressstrain}). The fracture propagates along weakly connected regions, indicating a brittle fracture mechanism.

For strain applied along the $y$-direction, the fracture onset occurs at a lower stress concentration level. At 9.5\% strain (Figure~\ref{fig:fracture}(d)), the stress remains well distributed, but at 11.5\% (Figure~\ref{fig:fracture}(e)), localized stress zones intensify in regions aligned with the periodic arrangement of the biphenylene units. The fracture develops in a more tortuous pattern, leading to partial reorganization of atomic bonds before the rupture occurs. The side view in Figure~\ref{fig:fracture}(e) shows signs of progressive bond stretching and delayed failure compared to the $x$-direction case.

\section{Conclusions}

In summary, we have proposed and characterized dodecaphenylyne (DP), a novel 2D carbon allotrope with an architecture composed of four-, six-, and twelve-membered rings. DFT calculations and machine learning-based interatomic potentials revealed that the structure is thermodynamically and dynamically stable, with a formation energy of –7.98 eV/atom. No imaginary phonon modes were observed, and structural integrity is maintained up to 1000 K. The system exhibits a moderate indirect band gap of approximately 1.73 eV (HSE06), characterized by asymmetric HOCO and LUCO orbitals and mixed sp/sp$^2$ hybridizations that significantly influence both its electronic and mechanical behavior.

The analyses of the DP mechanical properties revealed strong anisotropy, with Young's modulus values of 469 GPa and 600 GPa along the $x$- and $y$-directions, respectively. The ultimate tensile strength reached 52.9 GPa, and structural fracture is direction-dependent, as confirmed by von Mises stress maps. Carrier mobility calculations based on deformation potential theory yielded values as high as $30.6 \times 10^4$ cm$^2$/V$\cdot$s for electrons and $8.4 \times 10^4$ cm$^2$/V$\cdot$s for holes. DP demonstrated strong anisotropic optical absorption in both the visible and ultraviolet ranges, suggesting its suitability for optoelectronic and photonic devices. 

\begin{acknowledgement}

This work was supported by the Brazilian funding agencies Fundação de Amparo à Pesquisa do Estado de São Paulo - FAPESP (grant no. 2022/03959-6, 2022/00349- 2, 2022/14576-0, 2020/01144-0, 2024/05087-1, and 2022/16509-9), and National Council for Scientific and Technological Development - CNPq (grant no. 307213/2021–8). L.A.R.J. acknowledges the financial support from FAP-DF grants 00193.00001808/2022-71 and $00193-00001857/2023-95$, FAPDF-PRONEM grant 00193.00001247/2021-20, PDPG-FAPDF-CAPES Centro-Oeste 00193-00000867/2024-94, and CNPq grants $350176/2022-1$ and $167745/2023-9$. A.C.D acknowledges the following FAP-DF grants: 00193-00001817/2023-43 and 00193-00002073/2023-84, the following CNPq grants: 408144/2022-0, 305174/2023-1, 444069/2024-0 and 444431/2024-1, and the computational resources from Cenapad-SP (project 897) and Lobo Carneiro HPC (project 133). K.A.L.L. and D.S.G. acknowledge the Center for Computational Engineering \& Sciences (CCES) at Unicamp for financial support through the FAPESP/CEPID Grant 2013/08293-7. The authors gratefully acknowledge CAPES's financial support for publishing this work under the CC BY Open Access license through the ACS-CAPES agreement.

\end{acknowledgement}

\section*{Declaration of competing interest}
All authors declare no competing financial or non-financial interests.

\begin{suppinfo}
We provide the structural data of the proposed material in the CIF format.
\end{suppinfo}

\bibliography{refs}

\end{document}